\documentclass[10pt]{article}
\usepackage[all,knot,poly]{xy}
\xyoption{arc}
\usepackage{graphicx}
\usepackage{amssymb,amsmath}
\usepackage{epsfig}
\graphicspath{{./}{img/}}
\usepackage[latin1]{inputenc}
\usepackage{amsmath,amssymb, amsthm}
\usepackage{accents}
\usepackage{cancel}

\usepackage{enumerate}
\usepackage{fancyhdr}
\usepackage{eso-pic,graphicx}
\usepackage{pstricks}

\author{Enore Guadagnini}

\def\be{\begin{equation}}
\def\ee{\end{equation}}
\def\bea{\begin{eqnarray}}
\def\eea{\end{eqnarray}}
\def\ba{\begin{array}{rcl}}
\def\ea{\end{array}}

\catcode`@=11 
\font\sevensy=cmsy7
\catcode`@=12 
\newbox\novebox
\setbox\novebox=\hbox{\kern+.3em {$V_i$}%
\kern-0.68em\raise1.8ex\hbox{\sevensy {\char'016}}\kern+.3em} 

\hoffset = - 33 pt
\voffset  = - 5  pt
\def\tfract#1/#2{{\textstyle{\raise0.8pt\hbox{$\scriptstyle#1$}\over%
\hbox{\lower0.8pt\hbox{$\scriptstyle#2$}}}}}
\def\mezzo{\tfract 1/2 }
\def\imezzi{\tfract i/2 }
\begin{document}

\title{Functional integration and abelian link invariants}

\author{Enore Guadagnini}

\maketitle

\centerline{\sl Dipartimento di Fisica, Universit\`a di Pisa}
\centerline{\sl and INFN, Sezione di Pisa}

\vskip 2 truecm 

\begin{abstract}
The functional integral computation  of the various topological invariants,  which are associated with the Chern-Simons field theory,  is considered. The standard perturbative setting in quantum field theory is rewieved and new developments in the path-integral approach, based on the Deligne-Beilinson cohomology, are described in the case of the abelian $U(1)$ Chern-Simons field theory formulated in $S^1 \times S^2$.  

\end{abstract}

\vskip 9 truecm 

\noindent Talk given at the workshop {\sl Chern-Simons Gauge theory: 20 years after}, Bonn, August  2009     

\newpage

\baselineskip=14 true pt

\section{Introduction}

The main subject of my talk concerns the use of the so-called path-integral ---or functional  integration--- in the definition and in the computation  of the various topological invariants which are associated with the quantum Chern-Simons field theory.  This argument has already been introduced in several talks at this conference, so  I shall skip the preliminaries and I will concentrate on the following question, which has indirectly  been posed yesterday by one of the speakers. 

\smallskip

{\it We all agree  that the path-integral has not a precise  meaning. In particular, if $S_{CS}$ denotes the Chern-Simons action,  the functional integral}
\be 
 I(M)  = \int_M DA \;  e^{i S_{CS} [A] } \; , 
 \label{1}
\ee
{\it which should correspond to an invariant of the 3-manifold $M$,  is not well defined. So, what is the meaning of expression }(1) ?

\smallskip

\noindent  I shall try to present the answer to this question in simple but  rather complete terms. Some basic features of the use of the functional integration in quantum field theory will be described in section~2 and section~3. In section~4  some new developments in the path-integral computation of observables in Chern-Simons field theory will be presented.  I will show how to bypass the difficulties of standard perturbation theory in the case of the abelian $U(1)$ Chern-Simons theory formulated in a closed non-trivial 3-manifold $M$; the example $M = S^1 \times S^2$  will be discussed in detail.  A few observations on the 3-manifold  invariants associated with the Chern-Simons theory are contained in section~5 and section~6. 

\section{Perturbative quantum field theory} 

Functional integration can be used in perturbative quantum field theory. 
Given a set of fields ---denoted by $\phi (x)$--- and an action functional $S[\phi ]$, in physics one is usually interested in functional integrations with a quite peculiar ``measure"  which naively corresponds to the following product 
\be 
D \phi \; e^{i S [\phi ] }  =  
 \hbox {(const.)} \left (  \prod_x d \phi (x) \right ) \;  e^{i S [\phi ] } \; . \label{2}
\ee
Alternatively, one can introduce a complete set of orthonormal functions $\{ \psi_n (x) \}$, with $n \in {\mathbb N}$, so that  each  classical configuration $\phi (x) $ can  be written as a linear combination of the base functions, $\phi (x) = \sum_n c_n \psi_n (x)$. By varying the (real) coefficients $\{ c_n \}$ one gets an infinitesimal  variation $d \phi (x) $ of the fields, $d\phi(x) = \sum_n dc_n \psi_n (x) $. Then, one can replace expression (2) with 
\be
D \phi \; e^{i S [\phi ] }  =  \hbox {(const.)} \left ( \prod_{n=1}^\infty d c_n \right ) \; e^{i S [\phi  ] }  \; .
 \label{3}
\ee
It is well known that both expressions (2) and (3) are not well defined. With an infinite number of integration variables, the result of an integration looks like the product of an infinite number of real coefficients that, apart from very exceptional cases,  is not well defined. So, independently of the choice of the value of the multiplicative constant which appears in equation (2) or in equation (3), both expressions (2) and (3) do not represent a well defined integration measure. In facts, the integral of the measure $I = \int D \phi \, e^{iS} $, which can be imagined to represent some kind of ``partition function" ---precisely like expression (1)--- is in general not well defined. 

\subsection{Observables}

On the other hand, it is a fact that perturbative quantum field theory provides a rather accurate description of physical phenomena. For instance, in the  $SU(3)_c \times SU(2)_L \times U(1)_Y$ Standard Model, by means of the functional integration method one can compute the value of the magnetic moment of the electron (or of the other charged leptons). The prediction of the Standard Model can be compared with the observed experimental value of the magnetic moment of the electron.  The actual agreement \cite{XX} between the computed value and the real observed value is within less then one part in a million. In addition to the remarkable correspondence of the experiments with the predictions of quantum field theory, the puzzling question is: 

\medskip

{\it  How is it possible, by using a  ill-defined functional integration, to make a prediction ?} 

\medskip 

\noindent For, in order to make a prediction, no ambiguities must occur, all the steps of the computation have to be well defined and, independently of  any choice of notations or conventions,   the final expression/value of the prediction  must be unique.  

The point is that all the predictions of quantum field theory that can be compared with experiments ---quantities of this type are called ``observables"---  never take the form of the partition function $I = \int D \phi \, e^{i S} $. 
In standard quantum field theory, the observables are strictly connected with expectation values of the type 
\be
\langle F[\phi ] \rangle = \frac{\int D \phi \; e^{i S [\phi ] } \, F[\phi ]  }{\int D \phi \; e^{i S [\phi ] } } \; , 
\label{4}
\ee
where $F[\phi]$ is a functional of the fields. At first sight, expression (4) also appears to be  not well defined because, in order to compute $\langle F[\phi ] \rangle$, one could imagine to evaluate separately the numerator and  the denominator and  then to take the ratio. However, this is not what physicists do in order to compute the expectation values. In standard perturbative quantum field theory equation (4) means: choose some kind of  ``regularization" in order to give a meaning to the numerator and  to the denominator simultaneoulsy.  Then, for the  regularized ratio,  consider the limit in which the regularization is removed.  This limit exists, or it exists at least for the quantum field theory models that appear to be somehow related to the description of physical phenomena. I shall present one example in a while. The regularization of the path-integral measure must necessarily be expressed in terms of a finite number of integration variables. Thus, for example, one could regularize expression (3) by considering the finite product 
\be 
D \phi \; e^{i S [\phi ] }  \Bigr |_{\rm reg}  =  D \phi_{_N} \; e^{i S [\phi ] } = \hbox {(const.)} \left ( \prod_{n=1}^N d c_n \right ) \; e^{i S [\phi ] }  \; .
\label{5} 
\ee
In this case, expression (4) should be interpreted as
\be
\langle F[\phi ] \rangle =  \lim_{N\rightarrow \infty } \frac{\int  D \phi_{_N} \; e^{i S [\phi ] } \, F[\phi ]  }{\int D \phi_{_N} \; e^{i S [\phi ] } } \; .  
\label{6}
\ee

\noindent After this first logical settlement, the next step consists in disentangling the difficulties in the computation of $\langle F[\phi ] \rangle$  that are related to the form of the action $S[\phi ]$ from those that are connected with the structure of the functional $F[\phi ]$ itself. To this end, it is convenient to consider the so-called correlation functions 
\be
\langle \phi(x_1) \cdots \phi(x_n) \rangle = \frac{\int D \phi \; e^{i S [\phi ] } \, \phi(x_1) \cdots \phi(x_n)  }{\int D \phi \; e^{i S [\phi ] } } \; . 
\label{7}
\ee
(From now on it is understood that the meaning of a ratio of two functional integrals, like expression (7), is specified by a prescription of the type illustrated in equation (6).)
The correlation functions are determined by the form of the action $S[\phi ]$ and, provided they are well defined, one can then consider  the problem of expressing  $\langle F[\phi ] \rangle$ in terms of the correlation functions. In several field theory  applications in particle physics, this second task is rather trivial because from  the correlation functions one can  obtain directly the transition amplitudes for the various particles  processes. However, in the Chern-Simons field theory, the solution of this second problem presents peculiar non-trivial aspects. In the computation of the expectation values of the holonomies associated with oriented links, one has to introduce a framing procedure in order to eliminate the ambiguities ---which occur in the product of two $A$ fields at the same point---  and preserve the ambient isotopy invariance. 

In order to  control  the whole set of correlation functions, it is useful to introduce the generating functional  
\be
Z[J] = 
\frac{\int D \phi \; e^{i S [\phi ] } \, e^{i \int J \phi }  }{\int D \phi \; e^{i S [\phi ] } } \; ,     
\label{8}
\ee
where $J = J(x) $ is a classical   ``source" variable. 
Perturbative quantum field theory is based on the existence of the so-called ``free" fields. Hence, let us  illustrate the meaning of expression (8) in the simple case in which the action is a quadratic functional $S= S_0$ of the fields 
\be 
S_0[\phi ] = \mezzo \int dx \,  \phi(x) \nabla \phi (x) \; , 
\label{9}
\ee
where $\nabla $ is an appropriate differential operator\footnote{For instance, when the field model is used to describe one kind of   free spinless  particles, the operator $\nabla $ is given by $\nabla = - \eta^{\mu \nu} \partial^2 / \partial x^\mu \partial x^\nu - m^2 $, where $\eta^{\mu \nu }$ denotes the Minkowski metric.}.  
The following identity, which is not related to the path-integral at all, plays an important role. One has  
\be 
\ba 
e^{i S_0 [\phi ] } \, e^{i \int J \phi } &=&  \exp \left \{ \imezzi \int dx  \,  \left [ \phi (x) \nabla \phi (x) + 2 J (x) \phi (x) \right ] \right \} \\
&=&   \exp \left \{ \imezzi \int dx \, \widetilde \phi (x) \nabla \widetilde \phi (x)  \right \} \; \exp \left \{ - \imezzi \int dx \, dy \, 
J(x) \nabla^{-1} (x,y) J(y) \right \} \\ 
&=&  e^{i S_0 [ \, \widetilde \phi \, ] } \;  e^{-\imezzi \int J \nabla^{-1} J } \; , 
\ea
\ee
where 
\be 
\widetilde \phi (x) = \phi (x) + \int dy \, \nabla^{-1} (x,y) J(y) \equiv \phi (x) + \hbox {const. } \; , 
\label{11}
\ee
and $\nabla^{-1} (x,y) $ is a Green function for the $\nabla $ operator 
\be 
\nabla \cdot  \nabla^{-1} (x,y) = \delta (x-y) \; . 
\label{12} 
\ee
In general,  $\nabla^{-1} (x,y) $  satisfies certain analytic properties\footnote{The Green function $i \nabla^{-1}(x,y)$  is usually called the Feynman propagator and, in the case of free spinless particles,    it is given by  $i \nabla^{-1}(x,y)= i \int [d^4p / (2\pi)^4]\, e^{-ip (x-y)}\, (p^2 - m^2 + i \epsilon)^{-1}$.} 
which, in particle physics, must be consistent with some observed properties of the experimental data like  causality and energy positivity.  
 For the moment, let us assume
that $\nabla^{-1} (x,y) $ exists, I shall return to this point later. 
The identity (10) implies that the generating functional $Z_0 [J]$ for free fields can be written as 
\be 
Z_0[J] = e^{-\imezzi \int J \nabla^{-1} J }  \; \times \, 
\frac{\int D \phi \; e^{i S_0 [\phi + \hbox {\small const.}] }    }{\int D \phi \; e^{i S_0 [\phi ] } } \; .      
\label{13}
\ee
Now one finds the crucial point. In order to compute the ratio of the two functional integrations shown in equation (13), one must consider the limit (in which the regularization is removed) of regularized functional integrations. With a finite number of integration variables, the result of the integral is invariant under translation of these variables, thus  
\be 
\frac{\int D \phi \; e^{i S_0 [\phi + \hbox {\small const.}] }    }{\int D \phi \; e^{i S_0 [\phi ] } } \equiv \lim_{N\rightarrow \infty } \frac{\int  D \phi_{_N} \; e^{i S_0 [\phi + \hbox {\small const.}] }   }{\int D \phi_{_N} \; e^{i S_0 [\phi ] } } = 1 \; ,   
\label{14}
\ee
and the generating functional is then 
\be 
Z_0[J] =   \exp \left \{ - \imezzi \int dx \, dy \, 
J(x) \nabla^{-1} (x,y) J(y) \right \}   \; .      
\label{15}
\ee
Note that, even if the  computation  of $ Z_0[J] $ that has been presented here is somehow based on the functional integration method, in the whole argument no ill-defined functional integration has  been  really computed. $Z_0[J] $ is well defined and determines the value of all the observables of the free theory. For  instance,  the poles in the Fourier transform of $ \nabla^{-1} (x,y)$ fix the value of the particles mass. So, all the observables of the free theory do not depend at all on the value that one could imagine to give to the  partition function  $I_0 = \int D \phi \, e^{iS_0}$. This remains true also in the case of  interacting fields models, where the action contains cubic or quartic terms in powers of the fields.  In fact, perturbative quantum field theories can also  be formulated \cite{BO, IZ} without the introduction of  functional integration. 

\noindent {\bf Remark.} Equation (14) can also be interpreted as a {\it defining relation},  because equality (14) is precisely the only  property of the functional integration that is used in standard perturbative quantum field theory. 

To sum up, in the path-integral formulation of perturbative quantum field theories  there is really no need  of computing  the  partition function $I= \int D \phi \, e^{iS} $,  any functional integration of this type is not well defined\footnote{Something similar also happens in statistical mechanics where  the partition function $Z$,  which takes the form $Z= {\rm Tr }\,  e^{-H/kT}$, is not an observable. The observables are combinations of the normalized mean values or can be derived from the thermodynamic potentials in the thermodynamic  limit. For example, in order to determine the free energy of the system, one only needs to consider the leading term of the expansion of $\ln Z$ in powers of the volume (for instance)  in the thermodynamic limit.  As a result, if  one modifies the partition function and multiplies it by, say, a factor five,  $Z\rightarrow Z^\prime = 5 Z$,   the expression of the free energy does not change. Thus, $Z$  cannot be ---and in facts it is not--- an observable, whereas the free energy is.} and all the observables do not depend on it.  Clearly,  the fact that the functional integration $I= \int D \phi \, e^{iS} $ is not well defined is not connected with the possible  existence of an analytic continuation of the field model in the euclidean region (this subject is related to the analytic properties of the Feynman propagator). Also, as far as the observables of quantum field theory are concerned,  the question whether, in the functional integration,  one has to sum over smooth or singular configurations for the field variables  is a completely irrelevant issue.  

Finally, in analogy with the result of a gaussian integral with a finite number of integration variables, sometimes in literature one finds the relation 
\be 
I_0 = \int D \phi \, e^{iS_0} = \hbox{(const.)} \, {\hbox {Det}}^{-1/2} \left (  -i \nabla \right ) \; . 
\label{16}
\ee
Equation (16) is not a definition of the value of the partition function because the expression appearing on the r.h.s. (for the differential operators $\nabla $ that normally enter the action functional) is not well defined. Expression (16) can be used as a guess-suggesting reminder   for the properties of the regularized  functional integral.  For instance,  when   $\nabla $  smoothly depends on a parameter (or on a set of parameters)    $\lambda$,  from equation (16) one can guess  the expression for the logarithmic  variation of $I_0  $ with respect to $\lambda $, 
\be 
I_0^{-1}\, \frac {\partial I_0} {\partial \lambda } = - \mezzo \, \hbox{Tr} \left ( \nabla^{-1} \, \frac{\partial \nabla }{\partial \lambda } \right ) \; . 
\label {17}
\ee
Differently from equation (16), expression (17) is well defined and, in facts, its structure is similar to the structure of the correlation functions ---or of many of the observables--- in quantum field theory \cite{S}. 

In a ``free fields" model, the correlations functions can  be derived from expression (15),  
\be
\langle \phi(x_1) \cdots \phi(x_n) \rangle_{0} = \frac{\int D \phi \; e^{i S_0 [\phi ] } \, \phi(x_1) \cdots \phi(x_n)  }{\int D \phi \; e^{i S_0 [\phi ] } } =  \frac{ (-i)^n \, \delta^n \, Z_0 [ J] }{ \delta J(x_1)\cdots \delta J(x_n)}  \biggr |_{J=0} \; . 
\label{18}
\ee
If the functional $F[\phi]$ can be written as a smooth linear combination of the correlation functions, one can then evaluate the observable $\langle  F[\phi] \rangle$. Since the correlations functions are really distributions, the computation of $\langle  F[\phi] \rangle$ may present ambiguities when, for instance, a correlation function is integrated with a function that is not a test function or when,  in a correlation function,  one needs to consider the limit of two (or more) coincident points.  This problem, which is also present in an interacting fields model, is related (in part) to the so-called composite operators problem.  

\subsection{Interactions and renormalization}

In the case of an interacting fields model, the action $S[\phi ]$ is written as the sum of two terms,  $ S[\phi ] = S_0 [\phi ] + S_I[\phi ]$, where $S_0$ denotes the ``free" action  and $S_I$ contains the interaction terms.   The  generating functional $Z[J]$ of equation (8) can be written as 
\be
Z[J] = e^{iZ^c [J]} = 
\frac{\int D \phi \; e^{i S_0 } \, e^{i S_I } \, e^{i \int J \phi }  }{\int D \phi \; e^{i S_0 } } \Biggr / \, \frac{\int D \phi \; e^{i S_0 } \, e^{i S_I }   }{\int D \phi \; e^{i S_0 } } \; ,     
\label{19}
\ee
where the numerator 
\be
\langle e^{i S_I } \, e^{i \int J \phi } \rangle_0 = \frac{\int D \phi \; e^{i S_0 } \, e^{i S_I } \, e^{i \int J \phi }  }{\int D \phi \; e^{i S_0 } } 
\label{20}
\ee
denotes the sum of all the Feynman diagrams (which are constructed with the Feynman propagator $i \nabla^{-1}$,  the interaction vertices of $S_I$, and in which each external leg corresponds to $- \int dy \, \nabla^{-1} (x,y) J(y) $); and the denominator 
\be
\langle  e^{i S_I }\rangle_0 =  \frac{\int D \phi \; e^{i S_0 } \, e^{i S_I }   }{\int D \phi \; e^{i S_0 } }
 \label{21}
 \ee
just corresponds to the sum of the vacuum-to-vacuum diagrams, i.e. diagrams with no external legs. Actually, the sum of the vacuum-to-vacuum diagrams factorizes in the numerator and  cancels out with the denominator. So, there is no need of computing the vacuum-to-vacuum diagrams (which remain divergent even after the standard regularization/renormalization procedure has been introduced).  In conclusion,  in the derivation  of the correlation functions and of the observables of an interacting field theory, 
one never has to compute the value of the partition function $I  = \int D \phi \, e^{iS}$. 

A few remarks on the meaning of the renormalization in field theory are in order. By means of a Legendre transformation of  the functional $Z^c [J]$ of the connected correlation functions, 
 one can introduce the effective action  $\Gamma [ \varphi ] $, 
 \be 
\varphi (x) = \frac{\delta Z^c [ J] }{\delta J(x)} \quad , \quad \Gamma [\varphi ] = Z^c [J] - \int dx J(x) \varphi (x) \; ,  
\label{22}
\ee
which is the sum of the one-particle-irreducible Feynman diagrams in which the external legs are represented by the classical variable $\varphi $. The perturbative expansion of a generic (nontrivial) correlation function ---of the field theory defined by the action $S[\phi ]$--- is equal to the  perturbative expansion containing diagrams at the {\it tree level} only of a   new field model defined by the action that coincides with the functional $\Gamma [ \varphi ]$. Diagrams at the tree level contain no loops: so, they  present no ultraviolet divergences and maintain all the symmetries of the action. This means that, provided $\Gamma [\varphi ]$ is well defined, all the correlations functions are well defined. 
Since the effective action establishes how  the symmetries of the theory are realized and determines the values of the observables\footnote{The  magnetic moment of the electron, for instance, is described by the three-point proper vertex containing two spinor electron  fields  and one vector electromagnetic field. Namely, the renormalized effective action $\Gamma $ of the Standard Model admits an expansion in powers of the fields. Consider now the  term $B \in \Gamma $ given by 
$B = \int dx dy dz \, {\overline \psi_\alpha (x)} \Lambda^\mu_{ \alpha \beta} (x,y,z) \psi_\beta (y) A_\mu (z) $, where $\psi (x)$ is the spinor field associated with the electron and $A_\mu (x) $ denotes the 4-vector potential of electromagnetism. The  function $\Lambda^\mu_{ \alpha \beta} (x,y,z)$  (three-point  proper vertex) describes how the electron interacts with the electromagnetic field and contains, in particular, the required information on the magnetic moment of the electron.}, $\Gamma [\varphi ]$ is the fundamental functional that must be considered in the renormalization task.   

Quite often, some of the diagrams contributing to $\Gamma $ are not well defined and present ambiguities. The root of these ambiguities is usually due to the presence of divergences  which can be eliminated (in agreement with the action principle) by local conterterms, i.e. by terms which have  the form of integrals  of polynomials of the field variables (and  their derivatives) defined in the same point with divergent coefficients.  Generally, in the intermediate steps of the renormalization, one makes use of an arbitrary regularization and, after the introduction of appropriate local counterterms (which also depend on the choice of the regularization), one  takes  the limit in which the regularization is removed. 
The whole  renormalization procedure consists of: 

\begin{itemize}
 \item{}  introduction of local counterterms   (with divergent and finite coefficients) in the diagrams computations in order to make the effective action   $\Gamma $ finite  and maintain the maximum number of symmetries (Lorentz symmetry, internal and gauge symmetries,...);  

\item{} introduction of the normalization conditions, which determine the meaning of the finite physical parameters on which the renormalized $ \Gamma $ eventually depends. 

\end{itemize}

\smallskip 

In renormalizable models, only a finite number of parameters need to be fixed (coupling constants, particle masses, fields or wave functions normalizations). The  finite values of the coupling constants and of the particle masses  ---that must agree with the  experimental values--- are also  called the renormalized or physical parameters.  The coupling constants usually correspond to the values of certain transition amplitudes in particular kinematic conditions (for instance, in the limit of vanishing momenta), and the particle masses corresponds to the poles in the energy variable of the dressed propagators.  This means that, in the case of an interacting field theory,  a few specific terms of $\Gamma $ (and not of the action $S$)  determine the values of the renormalized or physical parameters. The renormalized parameters are observables, whereas the so-called bare parameters ---which enter the action functional $S[\phi ] $--- are not observables. 

The abelian Chern-Simons field theory in ${\mathbb R}^3$  is a ``free fields" theory because the action is a quadratic functional of the fields $S_{CS} = 2 \pi k \int A \wedge d A$; the renormalization is trivial in this case because the effective action coincides with the action. 

The Chern-Simons field theory in ${\mathbb R}^3$ with a simple non-abelian gauge group is an interacting theory and a non-trivial renormalization is required because  some of the contributions to the effective action have ambiguities. For instance, the one-loop correction to the two point function for the connection field $A$ is the sum of two terms: each term is divergent  but, in their sum, the two divergent parts tend to cancel. Consequently, this sum is not well defined  ($ \infty - \infty $ is not well defined) and has ambiguities. In agreement with the behaviour  of all  perturbative quantum field theories, these ambiguities take the form of a finite local term, namely $b \int A^a \wedge d A^a$, where $b$ is an arbitrary finite parameter. Since this term has the same structure of a lagrangian term, the ambiguity in the value of $b$ is totally irrelevant and produces no physical observable effects. (This kind of ambiguities, which is well known in quantum field theory, concerns the finite terms arbitrariness in the renormalization process.) In facts, any choice of the finite value of $b$ does not modify the structure of the proper vertices contained in  $\Gamma $  (it simply changes the name  of some bare unobservable parameter); consequently, the perturbative expansion of a generic correlation function ---which is equal to the perturbative expansion made of tree-level diagrams only, constructed with the functional $\Gamma $--- is not modified by a change of the value of $b$.
Thus, in the non-abelian Chern-Simons field theory in ${\mathbb R}^3$ with compact gauge group, the non-trivial aspect of the renormalization is concentrated in the normalization conditions, which state how to identify the coupling constant. In particular, consider  the complete (i.e. the sum of the contributions to all orders of perturbation theory) two-point proper vertex for the $A$ field  contained in $\Gamma $; its structure is fixed by the symmetries of the theory and  takes the form 
\be
\Gamma \; \Bigr |_{AA} = \alpha \,  \int A^a \wedge d A^a \; , 
\label{23} 
\ee
where the nonvanishing real coefficient $\alpha $ admits a power expansion in terms of the bare parameters. The dependence of $\alpha $ on the bare parameters is not unique and can be arbitrarily modified by  changing the regularization;  but how $\alpha $ depends on the bare parameters is not observable and then it is totally irrelevant. When the renormalized $\Gamma $ preserves the BRS symmetry, the normalization condition states that the coupling constant $k$ of the Chern-Simons theory is given by 
\be 
\alpha = k / 8 \pi \; . 
\label{24}
\ee
Like in any renormalizable field theory, all observables depend unambiguosly\footnote{For example, in the Standard Model the  magnetic moment of the electron depends unambiguosly \cite{XX} on the value of the renormalized electromagnetic coupling constant $\alpha_{em} \simeq 1/137$. Whereas the dependence of  $\alpha_{em}$  on the ``bare" coupling constant is not unique and is not observable; in facts, can you imagine how to measure it by means of an experiment?} on the renormalized parameters (in our case, on $k$). In particular, a  second order computation in perturbation theory of the expectation values of the holonomies shows that, in terms of the coupling constant $k$, the  deformation parameter $q$ turns out to be 
\be 
q = e^{- 2 \pi i / k } \; . 
\label{25}
\ee
The coupling constant of the non-abelian Chern-Simons field theory in three dimensions should not be confused with the so-called ``level" parameter which  appears in two-dimensional conformal field theories. 
 
 \section{Perturbative Chern-Simons field theory}
 
 Let us now concentrate on the Chern-Simons field theory \cite{W1} and on the path-integral computation of the  topological invariants. It is important to distinguish the cases in which the topological model is defined in ${\mathbb R}^3$  or in a closed 3-manifold $M$.  Also, it is significant to distinguish the expectation values $\langle W [A]\rangle $, of a gauge invariant functional $W [A]$ of the $A$ fields, from (a possible variant of) the  partition function  $I = \int DA \, e^{i S_{CS}}$.   
  
Let us firstly recall the results obtained in perturbation theory in ${\mathbb R}^3$.  The non-abelian Chern-Simons field theory formulated in ${\mathbb R}^3$  with compact gauge group is perturbatively renormalizable. Several  features of the perturbative expansion have been explored with the covariant gauge-fixing  of the so-called Landau gauge (explicit computations have been produced up to two loops). The primitive ultraviolet divergences associated with the single diagrams ---constructed with the connections and ghosts fields--- tend to cancel in the construction of the proper vertices. Actually, the theory is  finite \cite{FF} to all orders of perturbation theory. Therefore,  in the renormalization process,  only the  finite local conterterms freedom remains to be fixed by means of  the normalization conditions. These conditions determine how  to identify the coupling constant in the effective action.  

Since  renormalizability is determined by the short-distance behaviour of the model, the ultraviolet  properties  of the Chern-Simons theory formulated in ${\mathbb R}^3$ or formulated in  a generic 3-manifold $M$ are obviously the same. This does not imply that the use of perturbation theory extends trivially from ${\mathbb R}^3$ to $M$. There are in fact  obstacles ---for both the abelian and the non-abelian Chern-Simons theories formulated in a closed 3-manifold $M$---  to produce  real perturbative computations of the observables, but these difficulties are not related to the ultraviolet divergences. 

For the abelian Chern-Simons theory in ${\mathbb R}^3$,  the expectation values of the holonomies, which are associated with coloured oriented and framed links, can explicitly be computed in closed form by means of standard perturbation theory.  

In the non-abelian case with compact gauge group, the explicit computation  of the expectation values of  the Wilson line operators in ${\mathbb R}^3$  has been produced at the  third non-trivial order of perturbation theory \cite{HS}.  The results of perturbation theory are in agreement with what is expected on the basis of general arguments. In facts, by taking into account the relevant symmetry properties of the expectation values in quantum field theory ---namely, ambient isotopy invariance, validity of satellite relations, structure of the representation ring of the gauge group, covariance of the expectation values under a modification of the framing--- one finds that the expression of the expectation values (of the trace of the holonomies) is unique.  These invariants of framed and coloured links take the form of generalized Jones polynomials \cite{J}; the algebraic structure of these polynomials, which  is determined by  the characters of simple Lie groups, is very general. In facts,  these link invariants can also be obtained or defined by means of skein relations \cite{Q},  quantum group Hopf algebra methods \cite{T1}, statistical state models \cite{K1}. For each simple Lie algebra, the corresponding braid group representations entering the construction of these link polynomials have a universal and canonical structure (the classification of these braid group representations is somewhat  similar to the classification of the irreducible representations of simple Lie algebras). These braid group  representations also appear, for example,  as monodromy representations  in conformal bidimensional models \cite{C1, C2, F1}. 

The perturbative setting of the Chern-Simons field theory can be imagined to be extended from ${\mathbb R}^3$ to  a generic closed 3-manifold $M$. Indeed, provided  the fields  propagator is well defined, the whole perturbative construction  based on the Wick contractions trivially follows. However, as a matter of facts,  no explicit example  of a real functional integral computation ---in standard perturbation theory---  of an observable in a closed 3-manifold  $M \not= S^3$ has been produced in the last twenty years. 

\section{Functional integral and Deligne-Beilinson cohomology}

When the Chern-Simons field theory is formulated in a nontrivial closed oriented manifold $M\not= S^3$, the explicit computation of the observables by means of the standard perturbation theory within the path-integral  method presents technical difficulties, which are related to the gauge-fixing procedure and the definition of the fields propagator.  Let me show how it is  possible   to overcome these problems  in the abelian case \cite{EF}.    

\subsection{Basic definitions}

The abelian Chern-Simons theory \cite{ASS, ASS2, CRH}  with gauge group $U(1)$ is defined by means of a $U(1)$-connection A in a closed oriented 3-manifold $M$. The holonomy associated with an oriented knot $C\subset M$ is given by the integral $\int_C A$; this integral  is invariant under ordinary $U(1)$ gauge transformations acting on $A$. 
In the standard field theory formulation of abelian gauge theories, the configuration space locally coincides with the set of 1-forms modulo exact forms, $A \sim A + d \Lambda$. But if one assumes \cite{W1, FR} that a complete set of observables is given by the exponential of the holonomies $\{ \exp [  2 \pi i \int_C A ] \}$ which are associated with oriented knots  $C $  in $M$, the invariance group of the observables is actually  larger than the standard gauge group;  in facts,  the observables are locally defined on the classes of 1-forms modulo forms $\widehat A$ with integer periods, $A\sim A+ \widehat A  \; ,  \; \int_C {\widehat A} = n\in {\mathbb Z}$. More precisely, the configuration space is defined in terms of the Deligne-Beilinson (DB) cohomology classes \cite{D, B, FR}. 

The DB class associated with the connection $A$ will be denoted by the same symbol  $A \in H^1_D(M)$, where  $H^1_D(M)$ represents the DB cohomology group of $M$ of degree~1. Let   $H^3_D(M) $ be the space of the DB classes of degree~3; the  *-product of two classes is a pairing of the DB cohomology groups that defines a natural mapping \cite{PR} 
$$
H_D^1(M) \otimes H_D^1(M) \longrightarrow H^3_D(M) \; ;  
$$
the *-product of $A$ with $A$ just corresponds to the abelian Chern-Simons lagrangian 
\be 
A *A \longrightarrow A\, \wedge d A \; . 
\label{26}
\ee
Like the integral of any element of $H^3_D(M)$, the Chern-Simons action 
$$  
S_{CS} [A] = \int_M A* A  \longrightarrow \int_M A\wedge d A  
$$ 
is defined modulo integers; consequently, the path-integral phase factor 
$$ 
e^{ 2 \pi i k S_{CS}  [A] } =  e^{ 2 \pi i k  \int_M A * A }
$$ 
is well defined when the coupling constant $k$  takes integer values, $k \in {\mathbb Z}$, ($k \not= 0$). Let us now consider a framed, oriented and coloured link $L\subset M$ with $N$ components $\{ C_1 , C_2 , ... , C_N \} $.  The colour of each component $C_j$, with $j=1,2,...,N$, is represented by an integer charge $q_j \in {\mathbb Z}$.   The classical expression $W(L)$ of the Wilson line is given by 
\be
W(L) =  \prod_{j=1}^N \exp \Bigl \{ 2  \pi i q_j \int_{C_j} A  \Bigr \}
= \exp \Bigl \{ 2 \pi i \sum_j q_j \int_{C_j}  A \Bigr \} \; , 
\label{27}
\ee
and the observables of the Chern-Simons gauge theory in $M$ are given by the expectation values 
\begin{equation} 
\langle W(L) \rangle \Bigr |_{M} = \frac{\int_M DA \, e^{2 \pi i k S_{CS} [A] } \, W(L) }{\int_M DA \, e^{2 \pi i k S_{CS} [A] }} \; , 
\label{28}
\ee
where the path integral  should be defined on the DB classes  which belong to $H^1_D(M) $.  

\noindent {\bf Note.} One usually assumes that expression (28) is well defined.  But one should keep in mind that, for certain manifolds $M$ and for certain values of the coupling constant $k$, expression (28) could not be well defined.
 
\noindent The structure of the functional space admits a natural description in terms of the homology groups of $M$, as indicated by the following exact sequence \cite{S1, S2}
\be
0\buildrel \over \longrightarrow {\Omega ^1\left( {M } \right)}
\mathord{\left/ {\vphantom {{\Omega ^1\left( {M } \right)} {\Omega _{\mathbb Z}^1 \left( {M} \right)}}} \right. \kern-\nulldelimiterspace} {\Omega
_{\mathbb Z}^1 \left( {M } \right)}\buildrel \over \longrightarrow H_D^1
\left( {M} \right)\buildrel \over \longrightarrow H^2\left( {M} \right)\buildrel \over \longrightarrow 0 \; , 
\label{29}
\ee
where $\Omega^1( M )$ is the space of  $1$-forms on $M$, $\Omega _{\mathbb Z}^1 ( M )$ is the space of closed $1$-forms with integer periods on $M$ and $H^{p}( M )$ is the $( p)^{th}$ integral cohomology group of $M$. Thus, $H_D^1( M)$ can be understood as an affine bundle over $H^2(M)$, whose fibres have a typical underlying (infinite dimensional) vector space structure given by $\Omega^1(M) / \Omega^1_{\mathbb Z}(M)$.

\subsection{Distributional forms}

Now, the crucial observation is that, in order to compute the observables (28), the introduction of a gauge-fixing and of the fields propagator is not essential. In order to illustrate this point, let us consider first the case $M = S^3$. The integral  of a one-form $ A  $ along an oriented knot $C \subset S^3$ can be written as the integral on the whole $S^3$ of the external product $A \wedge J_C$, where the current $J_C$ is a distributional 2-form with support on the knot $C$; that is, one can  write   $\int_C A = \int_{S^3}A \wedge J_C$.  Since $C$ can be understood as the boundary of a Seifert surface $\Sigma_C$ in $S^3$, one has $J_C= d \eta_C $ for some $1$-form $\eta_C$ with support on $\Sigma_C$.  One then finds  $\int_C A = \int_{S^3} A \wedge d \eta_C$.  For example, consider the unknot $C$ in $S^3$ shown in  Figure~1,   with a simple disc as Seifert surface. Inside the open domain depicted in Figure~1, the oriented knot is described ---in local coordinates $(x,y,z)$--- by a piece of the $y$-axis and the corresponding distributional forms $J_C$ and $\eta_C $ are given by
$$
J_C = \delta (z) \, \delta (x) \, dz \wedge dx \qquad , \qquad \eta_C = \delta (z) \, \theta (-x)\, dz \; .
$$
In terms of  DB classes, one has 
\be 
\exp \Bigl \{ 2 \pi i \int_C   A  \Bigr \} = \exp \Bigl \{ 2 \pi i   \int_{S^3}  A * \eta_C \Bigr \} \; , 
\label{30}
\ee
where $\eta_C$ denotes the DB class ---associated with the knot $C$---  that is locally represented by the distributional form $\eta_C$.

\vskip 0.3 truecm 

\begin{figure}[htbp]
\centerline{\includegraphics[width=2.80in]{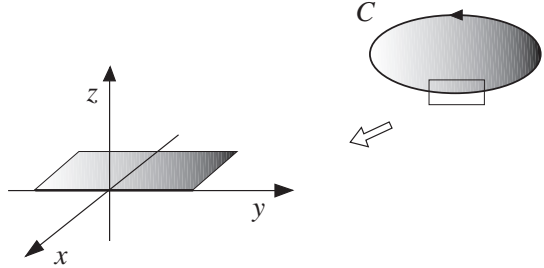}}
\caption{In a open domain with local coordinates $(x,y,z)$, a piece of the knot $C$  can be identified with the $y$ axis, and the disc  that it bounds can be identified with a portion of the half plane $(x<0,y,z=0)$.}
\end{figure}

\vskip 0.25 truecm 

\noindent For the coloured link $L\subset S^3$, one can write
\be
W(L) = \exp \Bigl \{ 2 \pi i \sum_j q_j \int_{C_j}  A \Bigr \} = 
\exp \Bigl \{ 2 \pi i \int _{S^3} A * \eta_L \Bigr \} \; , 
\label{31}
\ee
where $\eta_L = \sum_j q_j \eta_j $ denotes the DB class associated with the link $L$. 
  Since $H^2 (S^3)$ is trivial,  sequence (29) implies  $H_D^1( S^3) \simeq \Omega^1(S^3) / \Omega^1_{\mathbb Z}(S^3)$. 
The analogue of equation (10) now takes the form 
\be 
\ba 
e^{2 \pi i k  S_{CS} [A] } \, W(L) &=&  \exp \left \{ 2 \pi i k  \int_{S^3}  A * A  + 2 \pi i \int_{S^3} A* \eta_L  \right \} \\
&=&   \exp \left \{ 2 \pi i k \int_{S^3} \widetilde A * \widetilde A  \right \} \; \exp \left \{ - ( 2 \pi i / 4 k )   \int_{S^3}  \eta_L * \eta_L  \right \} \\ &=& e^{2 \pi i k  S_{CS} [\, \widetilde A \, ] } \,  \exp \left \{ - ( 2 \pi i / 4 k )   \int_{S^3}  \eta_L * \eta_L  \right \} 
 \; , 
\ea
\label{32}
\ee
where $\widetilde A = A - (1/2k) \, \eta_L$. The ambiguities in 
$\int_{S^3} \eta_L * \eta_L $ that are related to the self-linking number can be fixed in the standard way by the introduction of a framing for the link $L$. At this point, assuming invariance under translation of the functional integral (as shown in  equation (14)), one finally gets 
\begin{equation} 
\langle W(L) \rangle \Bigr |_{S^3} =  \exp \left \{ - ( 2 \pi i / 4 k )   \int_{S^3}  \eta_L * \eta_L  \right \} = \exp \Bigl \{ -( 2 i \pi / 4k)  \sum_{ij} q_i  {\mathbb L}_{ij} q_j  \Bigr  \} \; , 
\label{33}
\ee
 where ${\mathbb L}_{ij} $ are the matrix elements of the linking matrix associated with the link $L$. Equation (33) coincides with the result of standard perturbation theory but, in the derivation of expression (33),  both gauge-fixing and Feynman  propagator have not been introduced.  
 
The DB formalism turns out to be particularly useful for the path-integral computation of the observables in a generic 3-manifold $M$ because sequence (29) describes the non-trivial structure of the  functional space and equation (30) remains valid also in the case of a  knot $C $ in  a generic 3-manifold $M$. Indeed,  the class  $\eta_C \in  H_D^1 \left( M \right)$ which is canonically associated with a knot $C \subset M$ is  well defined for arbitrary 3-manifold $M$. 

 \subsection{Observables in non-trivial manifolds}
 
 As an example of functional integration in a non-trivial manifold, let us consider the Deligne-Beilinson formalism when the Chern-Simons theory is formulated in the manifold $M=  S^1 \times S^2$, which can be represented by the region of ${\mathbb R}^3$ which is delimited by two concentric 2-spheres, with the convention that  the points on the two surfaces with the same angular coordinates are identified. 
Since $H^2 (S^1 \times S^2) = {\mathbb Z}$, from  relation  (29) it follows that, as shown in Figure~2,  $H^1_D (S^1 \times S^2)$ can be understood as an affine bundle over ${\mathbb Z}$ in which each fibre has a linear space structure  isomorphic
to ${\Omega ^1\left( {S^1\times
S^2} \right)} \mathord{\left/ {\vphantom {{\Omega ^1\left( {S^1\times S^2}
\right)} {\Omega _{\mathbb Z}^1 \left( {S^1\times S^2} \right)}}} \right.
\kern-\nulldelimiterspace} {\Omega _{\mathbb Z}^1 \left( {S^1\times S^2}
\right)}$.

\vskip 0.25 truecm

\begin{figure}[htbp]
\centerline{\includegraphics[width=1.80in]{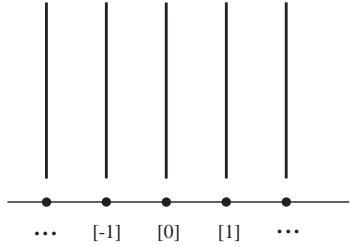}}
\caption{Presentation of the Deligne-Beilinson affine bundle
$H_D^1 \left( {S^1\times S^2} \right)$.}
\end{figure}

\vskip 0.18 truecm

\noindent In order to fix an origin in $H^1_D (S^1 \times S^2)$, we introduce the ``diagonal" section $s$,  
\begin{eqnarray}
\label{34}
 s : {\mathbb Z}  &\to&  H_D^1 \left( {S^1\times S^2} \right) \\
 n & \mapsto & s\left( n \right) \equiv  n  \,  {\gamma _0 }   \, ,  \nonumber
\end{eqnarray}
where $\gamma_0 \in   H_D^1 \left( {S^1\times S^2} \right)$ denotes the  DB class which is canonically  associated with the knot $G_0$, shown in Figure~3, that can be taken as a generator of $H_1(S^1 \times S^2) = {\mathbb Z}$. 

 \vskip 0.3 truecm

\begin{figure}[htbp]
\centerline{\includegraphics[width=1.1in]{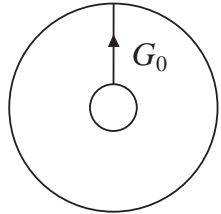}}
\caption{In the region of ${\mathbb R}^3$ that  provides a description of $S^1\times S^2$, the oriented  loop $G_0 \subset S^1 \times S^2$ ---generator of $H_1(S^1 \times S^2)$--- is  represented.}
\label{fig3}
\end{figure}

\vskip 0.12 truecm

\noindent Each element $ A \in  {H}_D^1 \left( {S^1\times S^2} \right) $ can then be written as
\begin{equation}
\label{35}
 A = n \,  {\gamma _0 } +  \alpha \; ,
\end{equation}
for some integer $n$ and  $ \alpha \in {\Omega ^1\left( {S^1\times S^2} \right)}
\mathord{\left/ {\vphantom {{\Omega ^1\left( {S^1\times S^2} \right)}
{\Omega _{\mathbb Z}^1 \left( {S^1\times S^2} \right)}}} \right.
\kern-\nulldelimiterspace} {\Omega _{\mathbb Z}^1 \left( {S^1\times S^2}
\right)}$. The (not regularized) functional measure takes the  form 
\be 
\ba 
  DA \, e^{2 \pi i k S_{CS} [A]} &=& \sum_{n= - \infty }^{+ \infty}  D  \alpha  \,
 \exp \left \{ 2 \pi i k \int_{S^1 \times S^2 }  \left(  n   \gamma _0  +  \alpha  \right) \ast  \left( n  \gamma _0 +  \alpha   \right)    \right\} \\
 &=& \sum_{n= - \infty }^{+ \infty}  D  \alpha  \,  \exp \left \{ 2\pi i k \int_{S^1 \times S^2 }    \alpha  \ast    \alpha   \right \} \,
 \exp \left \{ 4\pi i k \, n \int_{S^1 \times S^2 }     \alpha  \ast    \gamma _0 \right \} \,  .
\ea
\label{36}
\ee

Because of the non-trivial homology of the manifold $S^1 \times S^2$, the functional measure has a physical (not related to the gauge invariance) zero mode  $\beta_0 \in H^1_D(S^1 \times S^2) $. More precisely,  let us represent a generator of $H_2 (S^1\times S^2)$ by a oriented 2-dimensional sphere $S_0 $; $S_0$ is isotopic with the component $S^2$ of  $S^1 \times S^2$ and,  if one represents  $S^1 \times S^2$ by  the region of ${\mathbb R}^3$ which is delimited by two concentric spheres, $S_0 $ can  just be represented by a third concentric sphere, as shown in Figure~4.  

\vskip 0.20 truecm

\begin{figure}[htbp]
\centerline{\includegraphics[width=1.1in]{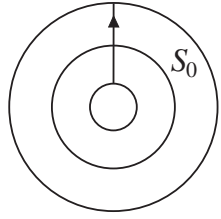}}
\caption{The sphere $S_0  \subset S^1\times S^2$.}
\label{fig4}
\end{figure}

\vskip 0.16 truecm

\noindent Let  $\beta_0$ be the distributional 1-form  which is globally defined in $S^1 \times S^2$ and has support on $S_0$, and let $\beta_0 \in H^1_D(S^1 \times S^2) $ be the class which is locally described by the distributional  1-form $\beta_0$. The overall  sign of  $\beta_0 $ is fixed by the orientation of $S_0$ so that
\begin{equation}
\int_{G_0} \beta_0 = 1 \, .
\label{37}
\end{equation}
Since the boundary of the closed surface $S_0$ is trivial, one has $ d \beta_0 = 0 $. From equation (37) and $ d \beta_0 = 0 $ it follows that, for any integer $m$, 
\be 
\exp \left \{  2 \pi i k S_{CS}[A] \right \} = \exp \left \{ 2 \pi i k S[A+  (m/2k) \beta_0 ] \right \} \; ,  
\label{38}
\ee 
that implies 
\be 
\exp \left \{  2 \pi i k S_{CS}[A] \right \} = \frac{1}{2k} \sum_{m=0}^{2k-1} \exp \left \{ 2 \pi i k S[A+  (m/2k) \beta_0 ] \right \} \; .   
\label{39}
\ee 
Consider now the observable (28) associated with the link $L \subset S^1 \times S^2$ for fixed integer $k$. The integral 
\be 
[L] = \int_{L} \beta_0 
\label{40}
\ee
 takes integer values; the value of $[L]$ is equal to the sum of the  intersection numbers (weighted with the values of the colour of the link components)  of the link $L$ with the surface $S_0$. Property (39) implies that $\langle W(L) \rangle \bigr |_{S^1 \times S^2}$ vanishes unless $[L] \equiv 0 $ mod $2k$  and, in that case, one can use again an identity of the type shown in equation (32)  to get  \cite{EF}
\be
\langle W(L) \rangle \Bigr |_{S^1 \times S^2} = 
 \left\{ \begin{gathered}
  0{\text{  if }} [L] \not\equiv 0{\text{ }}\bmod {\text{ }}2k \; , \hfill \\
  \exp \Bigl \{ -( 2 i \pi / 4k)  \sum_{ij} q_i  {\mathbb L}_{ij} q_j  \Bigr  \} {\text{  if }} [L] \equiv 0{\text{ }}\bmod {\text{ }}2k \; {\text{.}} \hfill \\
\end{gathered}  \right.
\label{41}
\ee
Expression (41) is the result of a real (non perturbative) functional integration computation in a non-trivial manifold. Equation (41) is in agreement with the prediction of the Reshetikhin-Turaev surgery rules, that will be discussed in the next section. 

The model considered in this section is an example of topological field theory  in which standard perturbation theory 
cannot be used.  In the quantum Chern-Simons field theory formulated in  $S^1 \times S^2$, the standard Feynman propagator for the $A$ field does not exist because of the normalizable zero mode that corresponds to the class $\beta_0 \in H^1_D(S^1 \times S^2) $. Among the field configurations,  one can find  a globally defined  1-form $ A_0$ such that $d A_0 = 0 $ but $A_0$ is not the gauge transformed of something  else.  

The Deligne-Beilinson formalism has been also applied \cite{EF}  to the torsion-free manifolds $M = S^1 \times \Sigma_g $ where $\Sigma_g $ is a closed 2-surface of genus $ g \geq 1$; and one example of manifold with torsion,   $M= RP^3$, has been studied by Thuillier in \cite{FT}. 

\section{Surgery invariants}

 In order to compute  ---by means of the quantum groups modular algebra--- topological invariants in a generic 3-manifold $M$,  Reshetikhin and Turaev  have produced  appropriate surgery rules  \cite{RT} that are in agreement with the  suggestions contained in \cite{W1}. 
 These rules  have been developed also  by Kohno \cite {K}, by Lickorish \cite{LIC} and by Morton and Strickland \cite{MOST}.  

According to the  Reshetikhin-Turaev surgery rules (adapted to 
the case of the abelian Chern-Simons field theory), for fixed integer $k$,   the expectation value of the Wilson line $W(L)$  associated with a link  $L $ in the 3-manifold $M= M_{\cal L}$ can be written as a ratio 
\be
\hbox{surgery~rules:}\qquad \langle W ( L) \rangle \Bigr |_{M_{\cal L} } =   \langle W ( L) \, W({\cal L}) \rangle  \Bigr |_{S^3}  \; {\Big  / } \;   \langle  W({\cal L}) \rangle \Bigr |_{S^3} 
\, .
\label{42}
\ee
In equation (42), ${\cal L}$ is the surgery link that codifies   a Dehn surgery \cite{ROL} presentation of $M$ in $S^3$; the integer surgery coefficients are determined by the framings of the components of ${\cal L}$. In the computation of the expectation values in $S^3$,  one has to sum over the values $q =0,1,2,..., 2k -1$ of the colours which are associated with the components of ${\cal L}$. The structure of equation (42) is somehow similar to the structure of the corresponding functional integral expression 
\begin{equation} 
\hbox{functional integral:}\qquad  \langle W(L) \rangle \Bigr |_{M} = \frac{\int_M DA \, e^{2 \pi i k S_{CS} [A] } \, W(L) }{\int_M DA \, e^{2 \pi i k S_{CS} [A] }} \; . 
\label{43}
\ee
But equation (42) is not based on a path-integral in the manifold $M$; the numerator and the denominator (which appear on the r.h.s of the equation)  ---that must be computed in $S^3$---  are separately well defined and, when the denominator is not vanishing, the ratio (42) is well defined.  Expression (42) refers to a particular surgery presentation described by the surgery link $\cal L$ but, since the ratio (42) is invariant under Kirby moves \cite{KI},  expression (42) is an isotopy invariant of the link $L$ in the oriented 3-manifold $M$.  

In all the examples considered so far, the computations of the expectation values $\langle W(L) \rangle \bigr |_{M}$ of equation (43),  which have been obtained by means of the  nonperturbative path-integral formalism based on the Deligne-Beilinson cohomology, are in agreement with expression (42).  

Assuming that equations (42) and (43) are equivalent, one finds that for certain manifolds and for particular integer values of the coupling constant $k$, the path-integral formula (43) is not well defined. For instance, in the case $M= RP^3$ the denominator of the ratio (42) is vanishing when $k$ is an odd integer and then expression (42) is not well defined. In the functional integral approach, the constraint on the values of $k$ when $M=RP^3$ has been discussed in \cite{FT}.

Equation  (42) implies  that \cite{EFM} : 

\begin{itemize}
 
 \item{} the set of expectation values of the  $U(1)$ Chern-Simons theory in $S^3$ and in any homology sphere $M_0$ coincide 
\be
\left \{ \langle W(L) \rangle \Bigr |_{M_0} \right \} = \left \{ \langle W(L) \rangle \Bigr |_{S^3} \right \} \; . 
\label{44}
\ee

\item{} For any coloured, oriented and framed link $L\subset M$, one can introduce a new link $\widetilde L\subset M$ 
(called the simplicial satellite of $L$) which is a satellite of $L$ and which is obtained from $L$ by replacing each link component of colour $q$ with $|q|$ parallel copies of the same component, each copy with unitary colour. If the simplicial satellite ${\widetilde L}  $ of the link $L  $  in a generic manifold $M$ is homologically trivial mod~$ 2k$, then  there exists a link  $L^\prime$ in $S^3$   such that   
\be
 \langle W(L) \rangle \Bigr |_{M}  =  \langle W(L^\prime) \rangle \Bigr |_{S^3}  \; . 
\label{45}
\ee

\end{itemize}

By means of the Reshetikhin-Turaev surgery rules, for fixed integer $k$, one can also define a 3-manifold invariant 
 \be 
I_k(M) = I_k(M_{\cal L} ) = \left ( 2k \right )^{- N_{\cal L}/2} \, e^{i\pi   \sigma ({\cal L}) /4} \langle W({\cal L} ) \rangle \Bigr |_{S^3}\; , 
\label{46}
\ee
where  $N_{\cal L}$ denotes the number of components of $\cal L$ and $\sigma ({\cal L})$ represents the so-called signature of the linking matrix associated with ${\cal L}$, i.e. $\sigma ({\cal L}) = n_+ - n_- $ where $n_{\pm }$ is the number of positive/negative eigenvalues of the linking matrix which is defined by the framed  link ${\cal L}$. Some properties of $I_k(M_{\cal L} )$ (and of its generalizations) have been studied, for instance,  in \cite{JP, FDE, HST}. If $M_0$ is a homology 3-sphere, then  \cite{EFM} one finds $I_k (M_0) = 1$.  One could suspect that the invariant $I_k(M)$ only depends on the homology group $H_1(M)$; the following  counterexamples show that this is not the case. The lens spaces $ L(5,1)$ and  $L(5,2) $  are not homeomorphic but they have the same homology group ${\mathbb Z}_5$;  equation (46) gives
\be 
I_2(L(5,1)) = -1 \qquad , \qquad I_2(L(5,2)) = 1\; . 
\label{47}
\ee

\noindent Similarly, the manifolds  $ L(9,1)$ and  $L(9,2) $  are not homeomorphic; they have the same homology group ${\mathbb Z}_9$ and are of the same homotopy type.  
One finds 
\be
I_3(L(9,1)) = i \sqrt 3 \qquad , \qquad I_3(L(9,2)) = - i \sqrt 3 \; . 
\label{48}
\ee

\section{Partition function}

In order to complete the answer to the question formulated in the Introduction, let us recall that,  in quantum field theory, any well defined functional integration takes really the form of a ratio of functional integrations. Therefore one can imagine that the   ``suitably normalized" partition function of the Chern-Simons theory formulated in the closed 3-manifold $M$,   
\be
R_{N_0}(M)  =  \frac{\int_{M} D A \; e^{i S_{CS} }    }{N_0 } \; ,  
\label{49}
\ee
should correspond to the  Reshetikhin-Turaev surgery invariant for the manifold $M$.   Equation (49) should be interpreted as the result of some limit prescription for the ratio of two regularized functional integrals, as indicated in equation (6).  So, $N_0$ stands for an appropriate path-integral that introduces  a  reference point for the integration.   Presumably,   $N_0$  is not unique; the specific choice of $N_0$ is precisely the crucial point  that will make expression (49) well defined.

Several variants of the Reshetikhin-Turaev surgery invariant have been introduced in literature and have been used to obtain well defined results; but all  these combinatorial invariants are not really based on a functional integration. We hope that, in the future,  a true functional integral derivation of an explicit  and well defined expression of a 3-manifold invariant will be produced.  

\bigskip 

\noindent {\bf Acknowledgments.}  I wish to thank F. Thuillier for useful discussions.

\end{document}